\documentclass[conference]{IEEEtran}
\IEEEoverridecommandlockouts%
\usepackage[utf8x]{inputenc}
\usepackage{cite}
\usepackage{textcomp}
\usepackage{amsmath,amssymb,amsfonts}
\usepackage{algorithmic}
\usepackage{graphicx}
\usepackage{textcomp}
\usepackage{xcolor}
\usepackage{booktabs}
\usepackage{url}

\usepackage{breakurl}
\usepackage[breaklinks]{hyperref}

\def\BibTeX{{\rm B\kern-.05em{\sc i\kern-.025em b}\kern-.08em
    T\kern-.1667em\lower.7ex\hbox{E}\kern-.125emX}}
\begin{document}

%-------------------- Otsikko ja nimet -------------------

\title{MEMS Heat Flux Sensor
\thanks{\copyright 2020 IEEE.  Personal use of this material is permitted.  Permission from IEEE must be obtained for all other uses, in any current or future media, including reprinting/republishing this material for advertising or promotional purposes, creating new collective works, for resale or redistribution to servers or lists, or reuse of any copyrighted component of this work in other works.}}
%Heat Flux Sensor using MEMS Manufacturing Methods
%Heat Flux Sensor with 3D MEMS structure
%MEMS-based manufacturing method for a Heat Pluq senser
%MEMS Heat Flux Sensor
%
\makeatletter%AT Letter = Anisotropic Thermoelement Letter
\newcommand{\linebreakand}{%
  \end{@IEEEauthorhalign}
  \hfill\mbox{}\par
  \mbox{}\hfill\begin{@IEEEauthorhalign}
}
\makeatother

\author{\IEEEauthorblockN{Antti Immonen}
\IEEEauthorblockA{\textit{LUT University}\\
Lappeenranta, Finland \\
antti.immonen@lut.fi}
\and
\IEEEauthorblockN{Saku Levikari}
\IEEEauthorblockA{\textit{LUT University}\\
Lappeenranta, Finland \\
saku.levikari@lut.fi}
\and
\IEEEauthorblockN{Feng  Gao}
\IEEEauthorblockA{\textit{VTT Technical Research}\\
\textit{Centre of Finland}\\
Espoo, Finland \\
feng.gao@vtt.fi}
\and
\IEEEauthorblockN{Mikko Kuisma}
\IEEEauthorblockA{\textit{LUT University}\\
Lappeenranta, Finland \\
mikko.kuisma@lut.fi}
\and
\IEEEauthorblockN{Pertti Silventoinen}
\IEEEauthorblockA{\textit{LUT University}\\
Lappeenranta, Finland \\
pertti.silventoinen@lut.fi}
}

\maketitle

% --------------- Dokumentti alkaa tästä ---------------------

\begin{abstract}
Heat flux sensors have potential in enabling applications that require tracking of direct and instantaneous thermal energy transfer. To facilitate their use, the sensors have to be robust and feasible to implement, while maintaining high sensitivity and a fast response time. However, most commercially available heat flux sensors are expensive to manufacture and have insufficient temporal responses. In this paper, a novel microelectromechanical heat flux sensor structure is proposed. The electrical performance of the prototype sensors is compared with commercially available heat flux sensors. Preliminary results show that our sensors have similar sensitivity and faster response compared to commercial sensors.
\end{abstract}
%Preliminary results show that the sensors have a comparable sensitivity and a faster temporal response as compared to their commercial counterparts.%comparison toistuu edelleen, pitää pohtia vielä uudelleen tätä.
%Ehdotuksia:
%Sensors have a comparable sensitivity with their commercial counterparts, and in addition possess a faster response time.
%The prototype sensors were found to have a faster response time and comparable sensitivity with their commercial counterparts.


\section{Introduction}
Heat flux sensors (HFS) are used in a growing number of applications requiring direct measurement of transferred thermal energy, where the use of commonplace temperature measurement is not practical or sufficient \cite{andrei, PracticalTempMeas}. %Application-specific advantages of heat flux sensors over conventional temperature measurements are typically related to the inherent properties of the sensor and the differential nature of the measurement, in other words, the operation of the sensor is based on heat conduction analysis across a relatively invariable medium inside the sensor\cite{PracticalTempMeas}. For one,  
Most notably, heat flux sensors enable direct measurement of the flow of heat even in the presence of changing heat coupling mechanisms, and in addition, facilitate measurement of submicrosecond heat transients \cite{andrei, yndrei}. The sensors can also be used for differential temperature measurement across the sensor itself, resolving differential temperatures on the order of microkelvins\cite{PracticalTempMeas, fSkin}.
 
Despite the potential benefits of heat flux measurement, HFSs are employed relatively rarely in the industry, and instead, the transfer rate of heat is calculated based on multiple temperature measurements \cite{andrei}. According to \cite{andrei}, the low popularity of discrete HFSs is most likely a result of the relatively high cost and challenging manufacturability, as well as low awareness of the capabilities of the technology. %Furthermore, the applicability of commercially available heat flux sensors and TEGs is often limited by the need for sophisticated instrumentation. Moreover, their response times  are often in the range hundreds of milliseconds \cite{fSkin}, which is insufficient for many applications.

%In measuring the heat transport properties of a system, a common requirement is to provvide minimal disturbance to the measurement \cite{AdvancesinHFMeas}. However, several typical heat flux sensor designs incorporate high thermal resistivity materials in forming the active layer in order to increase the voltage response of the device. This however disturbs the temperature field of the measured object, which may falsify the measurement result. Our sensor is constructed using high thermal conductivity materials, namely monocrystalline silicon, thermal conductivity of which typically has a value in excess of 100 W/mK. In comparison, a commercially available heat flux sensor by greenTEG AG has thermal conductivity of 1 to 1.5 W/mK. Thus, our sensor has the advantage of low thermal resistance and decreased disturbance of the measurement system's heat flux.

Recently, increasing attention has focused on employing thermoelectric transducers in IoT applications for both sensing and power generation purposes, which predicts an increasing demand for cost-effective and robust thermoelectric elements\cite{IoT_TE_review, IR_review, Y-typeHeatguideTEG}. An example of such an application is the use of thermoelectric elements in direct skin contact with the human body, where the sensor needs to have a high sensitivity while being mechanically durable \cite{Tshirt-TEG}. However, from the manufacturing and electromechanical viewpoints, such applications are not easy to implement with the currently available heat flux sensor technology\cite{andrei}. Thus, feasible and cost-effective manufacturing methods are needed. In the present paper, a  novel  3D microelectromechanical (MEMS) semiconductor sensor stucture is proposed. The preliminary findings related to an improved vertical 3D MEMS sensor structure % with enhanced exploitation of the sensing layer structures are presented. %millä tavalla eroaa ja parempi
and the performance of the sensor are presented.

\section{Sensor Design}
Discrete heat flux sensors are most commonly implemented as passive sensor elements that generate a voltage output in response to an imposed temperature gradient \cite{PracticalTempMeas}. While a number of widespread transducer technologies can be regarded as heat flux sensors, this paper focuses on contact-type thermopile sensors in the form factor of a plate, see Fig. \ref{fig:hfsenser}.

%planar 

Heat flux sensors are typically based on the Seebeck effect % because of the advantage of the sensor operating as a passive device,
which generates a voltage output \cite{andrei}. The voltage signal of a thermopile heat flux sensor results from a electromotive force induced along the length of an electric conductor by the thermal gradient applied. %Depending on the material properties, the direction of this induced voltage can be codirectional or opposite to the direction of the heat flow\cite{PracticalTempMeas}. 
A conventional thermopile is built by alternating two different types of thermoelectric legs, connected electrically in series and thermally in parallel, to sum up the generated voltages of individual legs into a useful amplitude\cite{PracticalTempMeas, IR_review}. The operation and main component parts of a traditional thermopile heat flux sensor are illustrated in Fig. \ref{fig:hfsenser}.

In addition to the classic plate-type heat flux sensors, heat transfer measurement can be achieved with several other transducer technologies based on the conversion of heat into electricity, including thermopile infrared (IR) sensors, Peltier elements, and thermoelectric generators (TEGs). While both TEGs and IR sensors are widespread in the industry and have low manufacturing costs\cite{IoT_TE_review, IR_review}, neither of the technologies is well-suited for surface heat flux measurements owing to their insufficient mechanical robustness or thermoelectric performance. 
\begin{figure}[h]
    \centering
    \includegraphics[width=0.9\columnwidth]{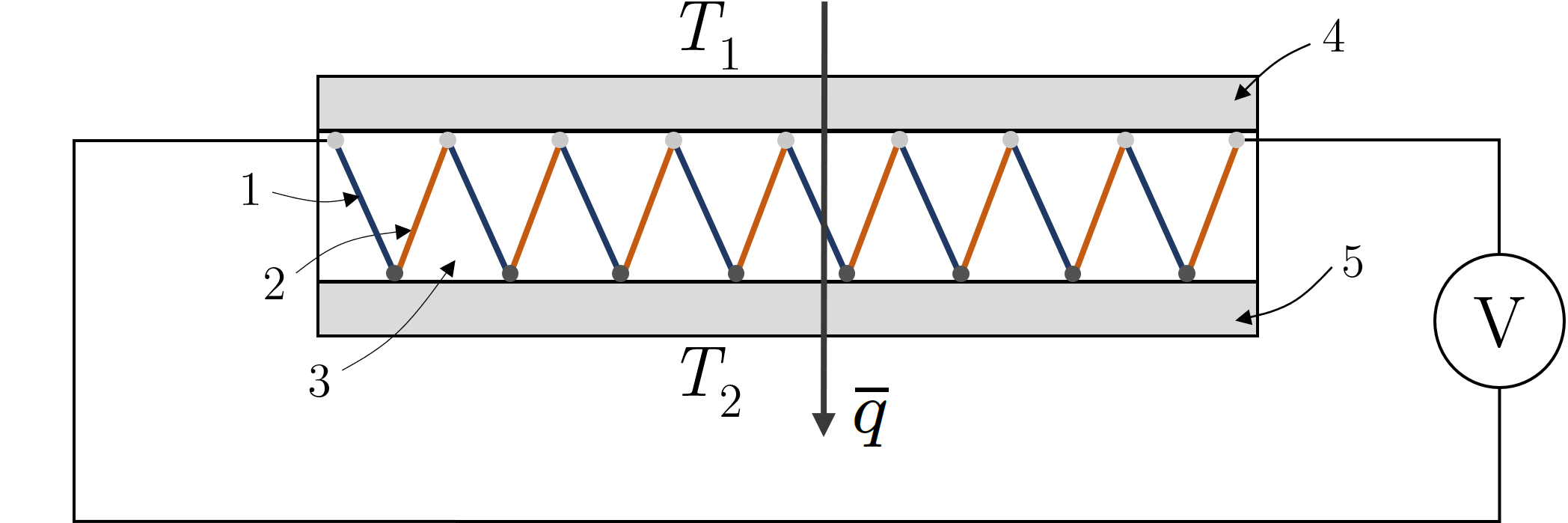}
    \caption{Thermopile HFS, composed of multiple thermocouple legs (1 and 2) in series. Voltage output corresponds to the heat flux $\overline{q}$ conducted through the heat resistance of the sensor active layer 3. In practical constructions, the sensors often include heat spreader support plates 4 and 5.}
    \label{fig:hfsenser}
\end{figure}
%Voltage is generated in alternating thermopile legs 1 and 2 in response to the heat flux $\overline{q}$ through sensor active layer 3. In practical constructions, the sensors often include heat spreader support plates 4 and 5.

Manufacturing a mechanically robust contact-type heat flux sensor with fast temporal response is difficult \cite{andrei}. An example of a mechanically durable sensor structure is the gradient heat flux sensor (GHFS), which incorporates an anisotropic active sensing layer based on a skewed multilayer structure \cite{yndrei}. Such a structure can be mechanically robust in a very wide range of environmental conditions without the need for applied protective cover plates, while achieving a significantly higher fill rate of the active layer volume than traditional thermopile structures \cite{andrei, yndrei}. However, sensors incorporating such structures are currently limited to low manufacturing volumes, as there are no commonly used mass production techniques available.

The use of MEMS manufacturing methods has been found to offer a number of benefits for achieving a high density of thermoelectric junctions in IR sensors\cite{IR_review}. However, owing to the lateral arrangement of the thermoelectric legs, the IR sensors have an uneven lateral heat distribution by design, which complicates their use in conductive heat flux measurements. Additionally, the suspended thin film devices are fragile \cite{huynh3D}, making them impractical for conductive heat flux measurement. To overcome these challenges, vertically configured 3D thermopile sensor structures have been proposed \cite{huynh3D, 2008PCB}. Such designs typically have a higher mechanical durability and a more uniform lateral heat distribution than conventional membrane-type sensors.%In the present paper, the preliminary findings related to an improved vertical 3D MEMS sensor structure with enhanced exploitation of the sensing layer structures are presented. 
%\subsection{Operation of the sensor}
 %The 98\% response time of the sensor can be calculated based on the conductive heat propagation through the sensor
%voltage gradient in the conductor in response to the thermal gradient

%\begin{equation}\label{eq:response time}
%\ t = \frac{1.5h^2}{\alpha} \,(\text{s}),
%\end{equation}
%where $h$ is the sensor thickness in meters and $\alpha$ is the thermal diffusivity in $\text{m}^2/\text{s}$ \cite{PracticalTempMeas}.

\subsection{Design optimization}
A common approach to improve the thermoelectric performance of a heat flux sensor is to employ materials with a high thermoelectric figure of merit zT\cite{IR_review}. The performance of a heat flux sensor can also be improved by optimizing the construction topology of the sensor \cite{IR_review, huynh3D}. To achieve an acceptable sensitivity while preserving a low series resistance, the number and packing density of the series-connected thermoelectric legs should be maximized. At the same time, optimal material properties should be maintained, that is, high conductivity and Seebeck coefficient, and low thermal conductivity\cite{PracticalTempMeas}. In addition, a high volumetric fill rate of the sensing layer with thermoelectrically contributing materials is desired\cite{yndrei}.
%as evident from the work of \cite{yndrei}, a high volumetric fill-rate of the sensing layer with thermoelectrically contributing materials is desired

%kuva Huynhin sensorista journal-lappuun+ sitten se Antin eka idea, jossa kunnon leikkauskuvat mikroskoopilla eri rakenteista
In \cite{huynh3D}, anisotropic wet etching was used to form a series of trenches, on top of which alternating metallization layers of Ni and Au were deposited to form a thermopile. The design removes the need for including both hot and cold junctions of the devices on the same plane, thus increasing the sensing area. Further improvements on the sensitivity of vertically configured MEMS devices may be attained by exploiting the structure forming the thermopile to a greater degree. The etched structures used to create a thermal resistance layer, such as the design of \cite{huynh3D}, can be incorporated as a part of the thermopile, when fabricated onto a SOI device layer. By making use of these structures, the density of thermoelectric legs can be increased, while simultaneously exploiting the higher Seebeck coefficient of silicon compared with that of metals. Further, employing thick crystalline silicon legs for the construction of the thermopile contributes to keeping the electrical resistance of the device low, and facilitates a fast response time.

\subsection{Sensor prototype manufactured in the study}
%To improve upon the existing heat flux sensor designs, a
A prototype batch of heat flux sensors was manufactured onto 6 inch SOI silicon wafers, Fig. \ref{fig:wafer}, with a 50 µm thick device layer. The process comprised of anisotropic etching of the single crystal SOI active layer to form an array of mesa structures and depositing and patterning a polysilicon and aluminum inter-layer on top of the mesa-structure to form a vertically configured 3D thermopile. In addition, the sensor surface was passivated with SiO$_2$ and the aluminum contact pads were revealed by etching a window through the passivation oxide. The structure of the developed sensor is illustrated in Fig. \ref{fig:sensorstructure}.
\begin{figure}[h]
    \centering
    \includegraphics[width=1\columnwidth]{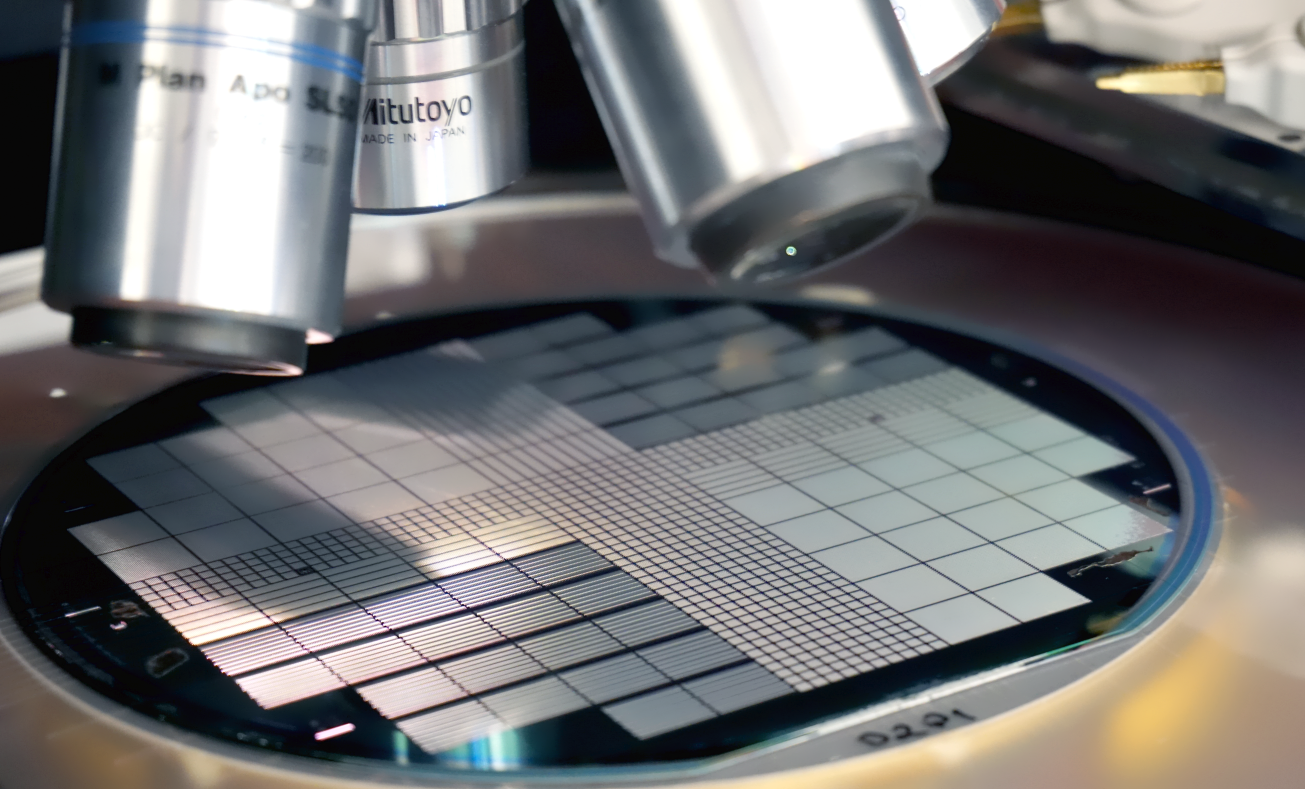}
    \caption{Processed silicon wafer in the test setup. The measurement system was built around a wafer probing station, with the wafer thermal chuck and the probe card holder (not shown) acting as thermal excitation source.}
    \label{fig:wafer}
\end{figure}
%Cross-sectional micrographs of heat flux sensors; a) commercial µTEG based on the conventional pillar structure, b) conventional MEMS IR thermopile sensor, c) Si-Al GHFS, and d) fabricated MEMS HFS prototype.
\begin{figure}[h]
    \centering
    \includegraphics[width=1\columnwidth]{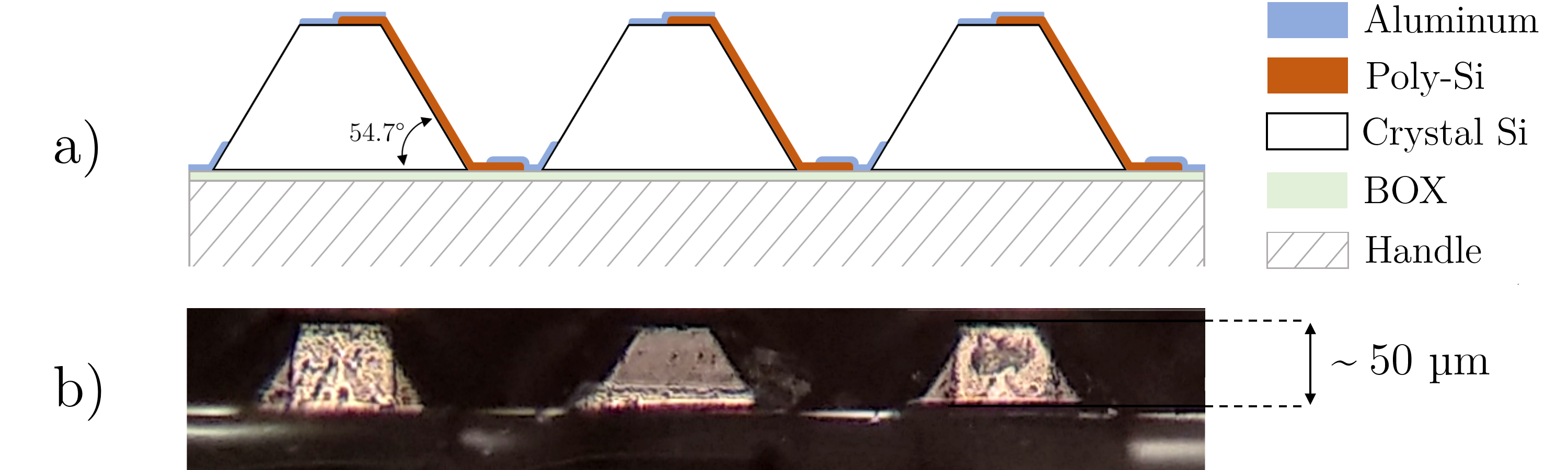}
    \caption{Cross-section of the fabricated sensor. a) Sensor structure, b) micrograpgh of the sensor.}
    \label{fig:sensorstructure}
\end{figure}

The manufactured sensor operates similarly to a conventional thermopile. The measured temperature gradient of interest is imposed across the SOI active layer, which incorporates electrically series-connected thermoelectric legs of p- and n-type silicon for transducing the signal. An array of mesa structures of n-doped single crystalline silicon are connected in series with p-type polysilicon films deposited on the slanting sidewalls of the mesa structures by a pattern of aluminum metallization. In such a configuration, in addition to forming the heat resistance layer, the device layer silicon bulk material contributes to the voltage response of the sensor.

% When a temperature gradient is imposed across the mesa-structures vertically, an electromotive force is generated based on the Seebeck effect in both the crystalline silicon bulk material and the polysilicon films.
\section{Measurements}
%early setup under construction
%test-setupin kuva?
Electrical measurements of the sensors were obtained during preliminary electrothermal characterization as described in \cite{DI}. %performed on four sensors as part of a test setup development project. The construction and operation of the test setup are described in \cite{DI}.
A stacked structure incorporating the device under test (DUT) and a calibrated commercial sensor around a wafer probing station was constructed, with the DUT sensors located adjacent to one another on the SOI wafer to ensure as equal thermal conditions as possible.

Relative characterization was performed by conducting a heat flux through the structure periodically in both directions during a 12 min test sequence. The reference heat flux was induced as a combination of conductive and radiative excitation by a combination of a thermal wafer chuck and a black body probe card assembly. During the excitation sequence, the transient and quasi-steady-state heat flux signals of both the DUTs and references, as well as several temperatures around the setting were recorded.

All the tested sensors produced a waveform similar to that of the reference sensor, with a slightly different transient response owing to differing thermal properties and contacts of the sensors. The  waveforms obtained from the test are shown in Fig. \ref{fig:result}.
\section{Results and Discussion}
%All the tested sensors produced a waveform similar to that of the reference sensor, with a slightly different transient response owing to differing thermal properties and contacts of the sensors. The  waveforms obtained from the test are shown in Fig. \ref{fig:result}.

Based on the preliminary measurements of the prototype sensors, the sensors were found to exhibit comparable sensitivities when compared with commercial sensors of similar size. The response time of the sensors was %estimated based on (\ref{eq:response time}) to be 
approximately 42 µs, which suggests a significantly faster response time as compared with the commercial sensors with response times of approximately 0.7 s\cite{fSkin}. In addition, a relatively low series resistance in the range of 400 to 700 ohm was measured. To further validate the operation of the proposed sensor, additional tests should be conducted, and new test batches fabricated.

Successfully implemented vertical sensor designs have the potential to ease the construction process of thermoelectric devices in both sensing and thermoelectric generator solutions. In cases where several current TEG designs have vertical-to-lateral heat guide structures and thermally and electrically insulating heater materials included, such as \cite{Y-typeHeatguideTEG}, the introduction of a successful vertical thermopile structure can significantly reduce the volumetric proportion of the filler materials and render the heat guide structures unnecessary. Moreover, if the same device structure is replicated on a smaller scale, the design has the potential for facilitating higher density of thermoelectric legs in direct contact with a measurement surface in comparison with the conventional lateral sensor designs.
%Lämmönjohtavuusasia
\begin{figure}[h]
    \centering
    \includegraphics[width=0.9\linewidth]{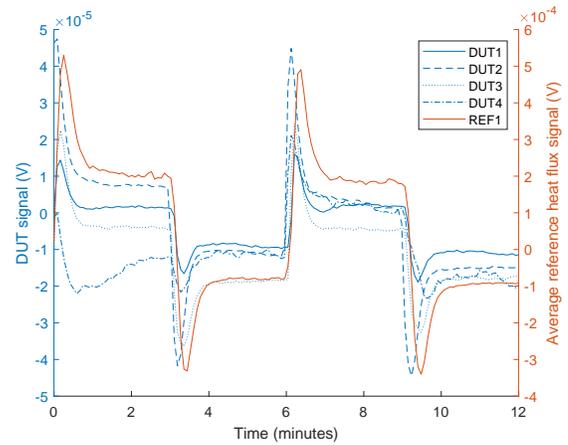}
    \caption{Output signals of the tested sensors (DUTs 1--4) recorded during a sequence of bidirectional thermal excitation, compared with a reference heat flux signal from calibrated sensor REF 1 placed under the DUTs.}
    \label{fig:result}
\end{figure}
\section{Conclusions}
A MEMS-compatible conductive heat flux sensor was successfully fabricated and tested. The promising early results %of relative characterization tests and calculations 
show that the developed sensors possess a sensitivity comparable with their commercially available sensors, approx. 1 Vmm²/W, and a faster response time of under 50 µs. The proposed fabrication method is compatible with MEMS processes, and thus facilitates cost-effective mass production of the sensors. Further, owing to the implemented vertical structure, the sensor is mechanically durable. These advantages facilitate the use of heat flux sensors in several new fields and applications, such as wearable electronics and IoT. % applications.
\section*{Acknowledgment}
The authors would like to thank greenTEG AG for providing the calibrated reference sensors. This work was supported by Business Finland through the project Q-Health.
%The authors would like to thank Feng Gao at the VTT Technical Research Centre of Finland Ltd for his much appreciated contribution to the project in the design and fabrication of the sensor prototype batch.

% --------------- Lähdeviitteet -----------------------

\bibliographystyle{IEEEtran}
\bibliography{references}

\end{document}